\documentclass[twocolumn]{aa}
\usepackage{graphicx}
\usepackage{amsmath}
\usepackage{amsfonts}
\usepackage{amssymb}
\usepackage{gensymb}
\usepackage{textcomp}
\usepackage[varg]{txfonts}
\usepackage[T1]{fontenc}
\usepackage{natbib}
\usepackage{epstopdf}
\usepackage{ulem}
\usepackage{booktabs}
\usepackage{hyperref}
\hypersetup{
    colorlinks,
    citecolor=blue,
    filecolor=blue,
    linkcolor=blue,
    urlcolor=blue,
    menucolor=black}
\newcommand{\RomanNumeralCaps}[1] {\MakeUppercase{\romannumeral #1}}

\makeatletter
\renewcommand*\aa@pageof{, page \thepage{} of \pageref*{LastPage}}
\makeatother

\begin{document} 

\title{Statistical investigation of decayless oscillations in small-scale coronal loops observed by Solar Orbiter/EUI}
\author{Arpit Kumar Shrivastav
      \inst{1,2,3},
      Vaibhav Pant\inst{1}
      \and
      David Berghmans\inst{4}
      \and 
      Andrei N. Zhukov \inst{4,5}
      \and
      Tom Van Doorsselaere \inst{6}
      \and
      Elena Petrova \inst{6}
      \and 
      Dipankar Banerjee \inst{1,2,7}
      \and
      Daye Lim \inst{4,6}
      \and
      Cis Verbeeck \inst{4}
      }

\institute{Aryabhatta Research Institute of Observational Sciences, Nainital, India-263002
     \and
         Indian Institute of Astrophysics, Bangalore, India-560034
    \and 
        Joint Astronomy Programme and Department of Physics, Indian Institute of Science, Bangalore 560012, India
    \and  
        Solar-Terrestrial Centre of Excellence - SIDC, Royal Observatory of Belgium, Ringlaan - 3 - Av. Circulaire, 1180 Brussels, Belgium.
    \and 
        Skobeltsyn Institute of Nuclear Physics, Moscow State University, 119991 Moscow, Russia.
    \and
        Centre for mathematical Plasma Astrophysics, Mathematics Department, KU Leuven, Celestijnenlaan 200B bus 2400, B-3001 Leuven, Belgium.
    \and
        Center of Excellence in Space Science, IISER Kolkata, Kolkata, India-700064
         }

   \date{Received **; accepted **}
  \abstract{ Decayless kink oscillations are omnipresent in the solar atmosphere, and they are a viable candidate for coronal heating. Although there have been extensive studies of decayless oscillations in coronal loops with lengths of a few hundred megameters, the properties of these oscillations in small-scale ($\sim$10 Mm) loops are yet to be explored. In this study, we present the properties of decayless oscillations in small loops embedded in the quiet corona and coronal holes. We use high-resolution observations from the Extreme Ultraviolet Imager on board Solar Orbiter with pixel scales of 210 km and a cadence of 5 s or better. We analysed 42 oscillations in coronal loops with loop lengths varying between 3 to 23 Mm. The average displacement amplitude is found to be 134 km. The oscillations period has a range of 28 to 272 s, and the velocity amplitudes range from 2.1 to 16.4 km s$^{-1}$.  {The variation in the loop length with the period does not indicate a significant correlation. The wave mode of these waves is uncertain, and standing waves are one possibility. Our results for the coronal seismology and energy flux estimates were obtained considering standing modes.} The observed kink speeds are lower than those observed in active region coronal loops. We obtain an average magnetic field value of 2.1 G. We estimated the energy flux with a broad range of 0.6-313 W m$^{-2}$.  Moreover, we note that short-period decayless oscillations are not prevalent in the quiet Sun and coronal holes. Our study suggests that decayless oscillations in small-scale coronal loops are unlikely to provide enough energy to heat the quiet Sun and accelerate solar wind in coronal holes.
 
  }
  \titlerunning{A Statistical Study of Kink Oscillations in Small-Scale Loops}
   \authorrunning{Shrivastav et al.}
   \maketitle
   \nolinenumbers
\section{Introduction}\label{intro}

The outer layers of the solar atmosphere maintain temperatures that are greater than those in the solar photosphere. To maintain the million-degree temperature of the solar corona, heating should balance the strong radiative losses. The energy flux needed to equipoise the energy losses in the quiet-Sun, coronal holes, and active region is $\sim$300 $\text{W m}^{-2}$, $\sim$800 $\text{W m}^{-2}$, and $\text{10}^4$ $\text{W m}^{-2}$ \citep{1977Withbroe}. One of the proposed heating mechanisms of energy transfer in the solar atmosphere is magnetohydrodynamic (MHD) waves \citep{2015Moortel, 2015Arregui, 2020Van_Doorsselaere, 2021Banerjee}. The transverse displacements of coronal structures in extreme-ultraviolet (EUV) images are interpreted as the standing \citep{Schrijver1999, Aschwanden1999} and propagating kink modes of MHD waves \citep{2011McIntosh,2014Thurgood, 2018Weberg, 2019Morton}. Since kink oscillations are ubiquitous in the solar atmosphere \citep{2015Anfinogentov}, they are candidates for coronal heating \citep{2020Van_Doorsselaere}.

 Standing kink oscillations have been categorized into two distinct regimes. The amplitude of the oscillations, that are excited by low coronal eruptions such as jets, flares, and coronal mass ejections \citep{Nakariakov1999, 2012White, 2016Goddard, 2016Sarkar}, decays with time, and they are defined as decaying oscillations. Many active-region loops have been observed to oscillate with no apparent decay in amplitude for a few periods. These are known as decayless oscillations \citep{2012Tian, 2012Wang, 2013Anfinogentov,2022Zhonga}.  The generation of decayless oscillations is the subject of ongoing debate, but evidence from modelling and numerical simulations suggests that these oscillations can be triggered by quasi-steady flows acting as footpoint drivers \citep{2016Nakariakov, 2020Karampelas}. Furthermore, the excitation of these oscillations due to periodic footpoint drivers has been explored in numerical simulations \citep{2017Karampelas,2019Karampelas, 2019Guo,2021Shi}. Recently, \cite{2021Karampelas} investigated the origin of decayless waves through vortex shedding, while \cite{2020Afanasyev} demonstrated that the observational signature of decayless oscillations could also be reproduced by random footpoint driving. The decayless kink oscillations have been studied for active regions \citep{2013Anfinogentov, 2013Nistic, 2015Anfinogentov, 2019Anfinogento} and quiescent loops \citep{2018Duckenfield} (for a detailed review on kink oscillations, see \citealt{2021NakariakovSSRv}).

\begin{table*}[!ht]
\caption{Datasets used in this study.}  
\label{table1}
\begin{center}
\centering
   \begin{tabular} {@{}cccccccc@{}}

         \hline

  \rule{0pt}{3ex} Date & Time Interval of & Distance from  & Stonyhurst heliographic  & Plate & Cadence  & Field of   \\
  \rule{0pt}{3ex}  & observation (UT)  & the Sun (a.u) & longitude (deg) & scale (km) & (s) &  view (Mm$^{2}$)   \\

     \hline
      \rule{0pt}{3ex} 2021-09-14  & 04:08 - 04:26 & 0.59 & -47 & 210 & 5  & 430$\times$430 \\ 
       \rule{0pt}{3ex} 2021-09-14  & 05:53 - 06:11 & 0.59 & -47 & 210 & 5 & 430$\times$430 \\
  
       \rule{0pt}{3ex} 2022-03-30  & 04:30 - 04:59 & 0.33 & 93 & 119 & 3 & 244$\times$244  \\

  \hline
\end{tabular}
\end{center}
\end{table*}
\begin{figure*}[!ht]
\centering
\includegraphics[width=0.98\textwidth,clip,trim=0cm 1.9cm 0cm 0.82cm]{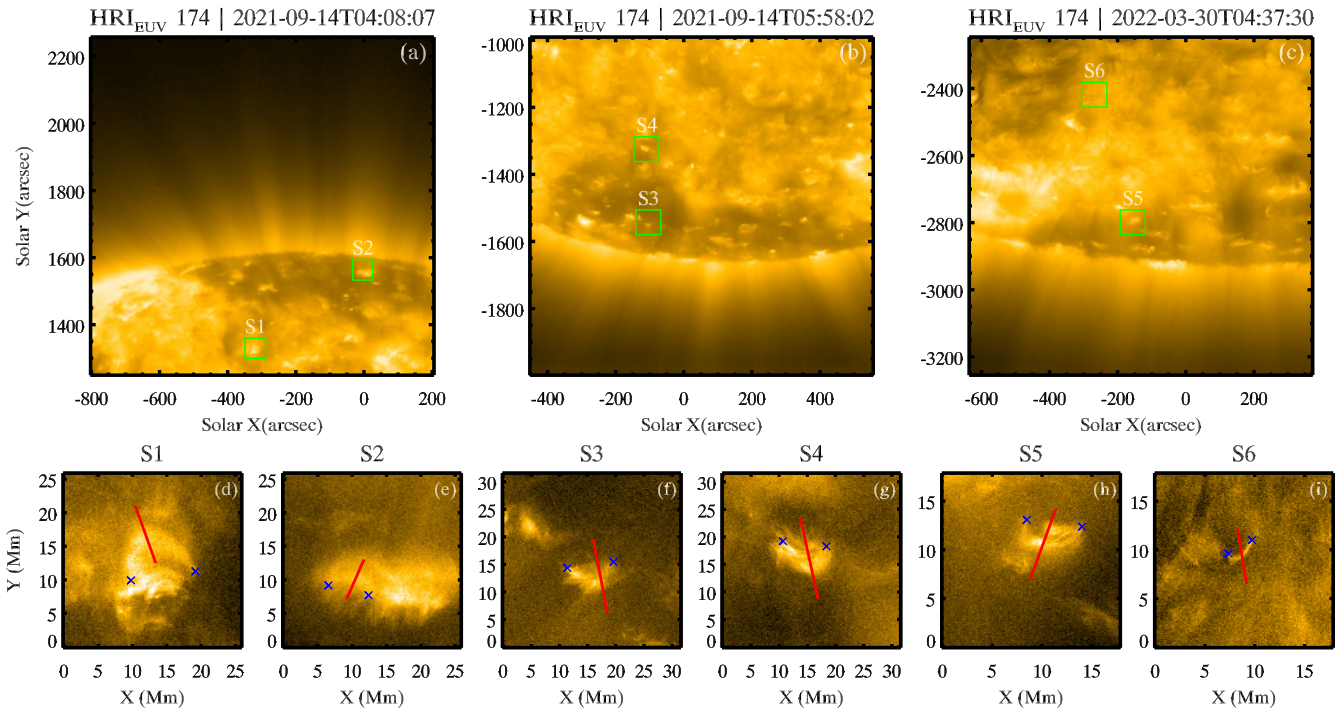}
\caption{Description of events. Panels (a), (b), and (c) represent the context images of the three datasets. The green boxes show examples of the selected loops for which oscillations are detected. The other loops we studied appear at different times and are not shown here. A magnified view of the loops in the green boxes for each upper panel is shown in the lower panels. The red lines in panels (d)-(i) depict the position of the artificial slits we used to generate the $x-t$ maps. The blue crosses in each panel show the approximate position of the footpoints. Details of the dataset are provided in Table \ref{table1}. An animation related to this figure is accessible \href{https://drive.google.com/file/d/1N-l4Mkp_-WMO9em_rplFFH7kam8zMGuC/view?usp=sharing}{online}}
\label{fig:context_sltpos}
\end{figure*}

In the past, several attempts have been made to investigate whether transverse waves carry sufficient energy to heat different regions of the solar atmosphere. Several studies have reported transverse waves in the solar atmosphere and estimated the energy fluxes by measuring the non-thermal line broadening of the transition region and coronal emission lines \citep{1990Hassler, 1998Banerjee, 2013Hahn}. These authors advocated that the energy fluxes estimated using the non-thermal line widths are sufficient to heat the solar corona and accelerate solar wind \citep{1990Hassler, 1998Banerjee, 2013Hahn}. Doppler velocity fluctuations, which are direct signatures of transverse waves, observed from the Coronal Multi-Channel Polarimeter \citep[CoMP; ][]{2008Tomczyk},  were found to be surprisingly small, which indicated that these waves do not provide sufficient energy to heat the active regions \citep{2007Tomczyk}. It was later found that the large line-of-sight integration in the CoMP data leads to the underestimation of the wave amplitudes that were estimated using Doppler velocity fluctuations \citep{2012McIntosh, 2019Pant}.

The observations of the chromosphere and transition region reveal that kink waves with significant energies to support the energy losses are omnipresent \citep{2007De_Pontieu, 2012Morton, 2014Tian}.  The High Resolution Coronal Imager \citep[Hi-C; ][]{2014Kobayashi_hic} observations of active regions indicated that coronal loops contain only little energy that might support heating \citep{2013Morton}. The study of transverse motions in fine-scale structures in active region moss revealed that kink waves have a higher energy in the lower layers of the atmosphere than in the corona \citep{2014Morton}. These fine-scale structures were not well resolved by the Atmospheric Imaging  Assembly (AIA; \citealt{Lemen2012}) on board the Solar Dynamic Observatory (SDO; \citealt{SDO2012}). MHD waves play an important role in accelerating the solar wind. The large-scale structures in coronal holes have been explored widely in this context and are seen to exhibit transverse motions \citep{2011McIntosh, 2014Thurgood, 2015Morton_natcom,2019Morton,2020Weberg}. The energy flux of kink waves in open field structures in coronal holes is computed to be lower than required for solar wind acceleration \citep{2014Thurgood, 2015Morton_natcom, 2018Weberg}.

Transverse standing waves have been analysed in quiet-Sun loops as well. {\cite{2018Duckenfield} found two periods in decayless oscillation in a large quiet-Sun loop.}  The quiet Sun and coronal holes, when seen in X-ray and EUV images, are permeated by small-scale loops and coronal bright points (CBPs) \citep{1976Golub, 2015Alipour}. Recently,  \citet{2022Gao} analysed decayless kink oscillations in CBPs using AIA. The study indicated that decayless oscillations are common in CBPs. The average loop length in their study was $\sim$23 Mm. The resolution of AIA does not allow us to detect low-amplitude transverse oscillations in CBPs, and therefore, the motion magnification technique was applied to enhance the oscillations. \cite{2022Gao} showed that the energy flux in kink waves in CBPs is not enough to support the radiative losses.

 The high resolution and cadence of the Extreme Ultraviolet Imager \citep[EUI;][]{2020A&A...642A...8R} on board Solar Orbiter \citep{2020A&A...642A...1M} has enabled us to probe the dynamics of small-scale coronal loops, whose signatures in AIA are largely unclear \citep{2021MandalSudip}. \cite{2022Petrova} performed a case study of two short-period decayless oscillations in the quiet-Sun region using EUI. The estimated loop lengths were $\sim$5-10 Mm.  Although these oscillations were observed in the quiet Sun, the energy flux of the wave was calculated to be approximately of a magnitude to balance even active region energy losses. \cite{2022Dong} conducted a statistical analysis of 111 small-scale active region loops with an average length of $\sim$15 Mm. The analysis identified short periods of oscillations ranging from $\sim$11 to 185 s, with energy fluxes spanning from $\sim$7 to 9220 W m$^{-2}$. The findings of \cite{2022Dong} revealed that the median wave energy flux of these oscillations is considerably lower than the energy required to heat the active region corona. Additionally, the authors found a strong correlation between the loop length and period. Although several studies of transverse oscillations in the polar region for large-scale structures exist, the kink waves in small loops rooted in coronal holes and in the quiet Sun have not been examined extensively.  We present a statistical investigation of the oscillation properties of small coronal loops (of $\sim$ 10 Mm) that are rooted in the coronal holes and quiet regions of the solar corona using high-resolution observations from EUI. The paper is structured as follows: Section \ref{obs} presents the datasets. Section \ref{analysis} describes the technique we used. The results and discussions are presented in Section \ref{results}, followed by Section \ref{conclusion}, which concludes the work.

\begin{figure*}[!ht]
\centering
\includegraphics[width=0.9\textwidth,clip,trim=0cm 0.5cm 0cm 0cm]{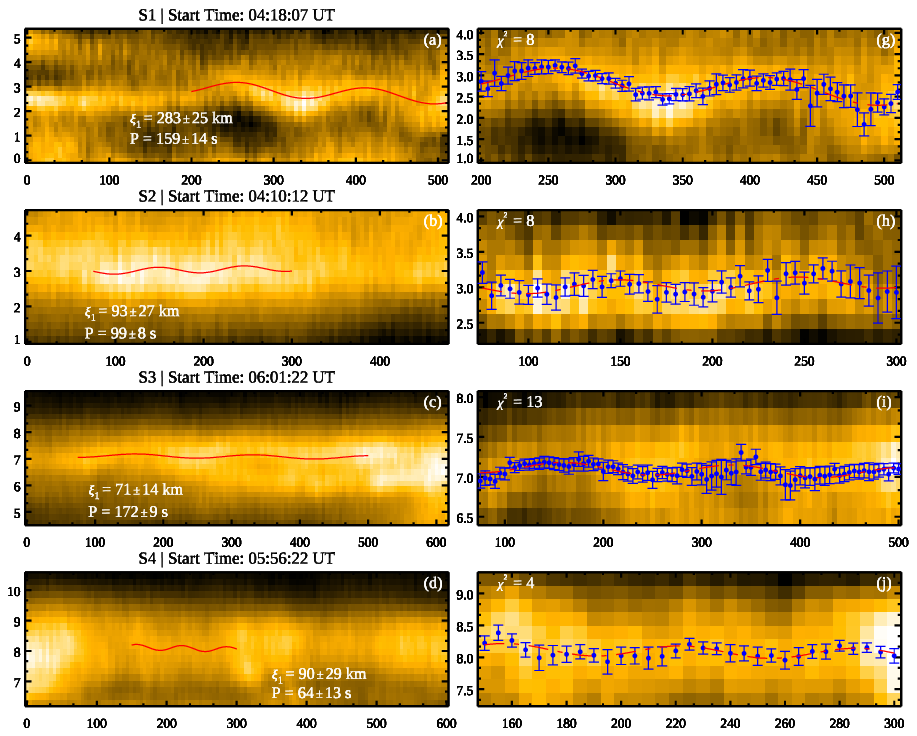}
\includegraphics[width=0.9\textwidth,clip,trim=0cm 3.9cm 0cm 0.1cm]{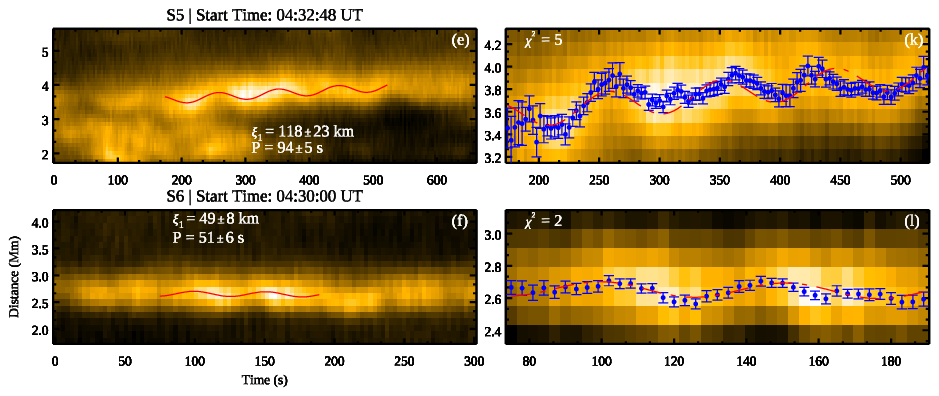}
\caption{Overview of $x-t$ maps. Panels (a)-(f) show the $x-t$ maps produced for slits S1 to S6 indicated in Figure \ref{fig:context_sltpos}(d)-(i). The red curves depict the best fit for the oscillations. {The error bars represent standard errors on the centre position of the loop.} The amplitude, $\xi_{1}$, and period, P, of the oscillation, along with the propagated errors, are written close to the fitted oscillations. A few $x-t$ maps show only part of the slits so that the oscillations are better visible. Panels (g)-(l) show the fitted oscillations along with the error bars for the position of the loop. These correspond to the $x-t$ maps shown in panels (a)-(f).  The dashed curves show the overplotted best-fit model. The chi-square values for the fits are also provided.}
\label{fig:xt-maps}
\end{figure*}

\section{Observation and data}\label{obs}

\begin{figure*}[!ht]
\centering
\includegraphics[width=0.8\textwidth,clip,trim=0cm 0.15cm 0cm 0.35cm]{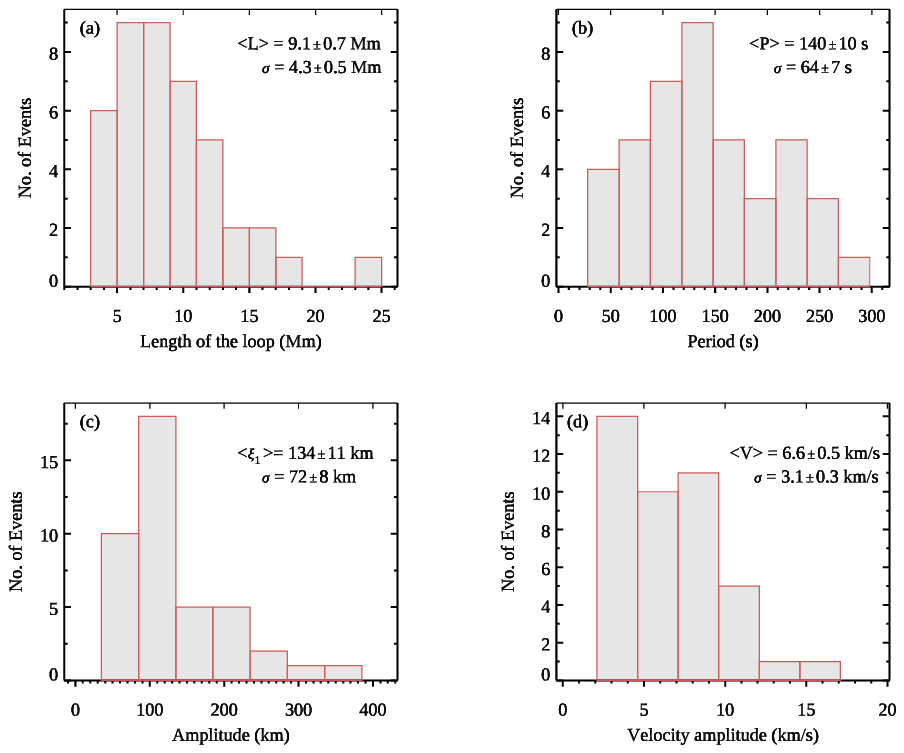}
\caption{Distribution of the oscillation parameters. The estimated parameter values are displayed as histograms, showing the distribution of (a) the loop length, (b) the period, (c) the displacement amplitude, and (d) the velocity amplitude. These histograms illustrate the range of values for each parameter. The average and standard deviation are provided in the respective panels. The plots also include the standard error in the mean and the standard deviation.}
\label{fig:dist_param}
\end{figure*}

The observational datasets we used in this study were obtained from the High Resolution Imager (HRI$_{\rm{EUV}}$) telescope of EUI on board Solar Orbiter. HRI$_{\rm{EUV}}$ observes the corona at 174 \AA\, which is attributed to Fe \RomanNumeralCaps{9} and Fe \RomanNumeralCaps{10} ions, and it captures the coronal dynamics of plasma at a temperature of $\sim$ 1MK. It has a field of view (FOV) of $\sim$ 17\arcmin$\times$ 17\arcmin with a plate scale of 0.492\arcsec pixel$^{-1}$.

The level 2 images were obtained from the EUI data release 5.0 \citep{euidatarelease5} and 4.0 \citep{euidatarelease4}. The EUI images are affected by spacecraft jitter, which we corrected for, and the images were aligned by applying the cross-correlation technique, as suggested in \cite{2022Mandal}.     

Figure \ref{fig:context_sltpos}(a)-(c) shows snapshots of the FOV of the Sun observed by HRI$_{\rm{EUV}}$ for three different datasets. These snapshots show many small-scale loops.  The green boxes outline the loops for which oscillations are observed. These images also show large-scale polar coronal holes and plumes. The datasets are focused on the quiet Sun and polar regions, with cadences suitable for observing short-period kink oscillations. Details of the datasets we used to identify the loops and build the statistics are described in Table~\ref{table1}. 

We selected several small-scale loops for our study and found 42 oscillating events in 40 coronal loops. These small-scale loops appear to be dynamic when investigated using animation, and many loops appear and disappear during the time interval of the datasets.  The properties of some of these loops are similar to active region fine-scale loops because they possess drifting motions \citep{2022Dong}.

\section{Analysis}\label{analysis}

Figure \ref{fig:context_sltpos}(d)-(i) shows the small-scale loops outlined by green boxes corresponding to the upper panels (a), (b), and (c). We placed slits approximately perpendicular to the loop axis. They are shown by red lines in each panel. These slits are 5 pixels wide and were used to generate distance-time ($x-t$) maps. The intensity was averaged over the width of the slits to increase the signal-to-noise ratio \citep{2012White&Verwichte, 2013Nistic}.

Figure \ref{fig:xt-maps} shows the $x-t$ maps of slits S1 to S6 as displayed in Figure \ref{fig:context_sltpos}(d)-(i). The $x-t$ maps reveal transverse oscillations. These $x-t$ maps indicate only a part of the slit to show the oscillations better.  The noise in relatively dimly illuminated areas such as coronal holes and the quiet Sun (Figure \ref{fig:context_sltpos}) is dominated by the sensor read-out noise, in particular, the photon shot noise. This can be seen as salt-and-pepper noise in the subfields in Figure \ref{fig:context_sltpos}(d)-(i). In order to optimise the telemetry, the images were Poisson recoded \citep{2005Nicula}
and compressed by a  JP2000-based compression scheme \citep[WICOM,][]{2020A&A...642A...8R} before transmission. During high-cadence HRI$_{\rm{EUV}}$ observations, this process is commanded in a high-quality mode such that artefacts are entirely insignificant. The absence of obvious horizontal streaks in Figure \ref{fig:xt-maps}, other than the expected solar features under study, also confirms that possibly remaining fixed noise patterns are negligible after the dark and flat corrections. We thus approximated the uncertainty in the intensity, I, in DN, using the following formula \citep{2022Petrova}
\begin{equation}
    \sigma_{\text{DN}}^{2} = \sigma_{\text{readout}}^{2}+\sigma_{\text{photon}}^{2}.
\end{equation}
$\sigma_{\text{readout}}$ denotes the readout noise of the HRI$_{\rm{EUV}}$ detector, estimated to be 2 DN. Additionally, the variance in the photon noise($\sigma_{\text{photon}}^{2}$) is determined by the product of the gain factor, g (with a value of 6.85 DN/photon, \citealt{2022Petrova}), and the intensity, I. Subsequently, the obtained uncertainty values were employed as errors in intensities to fit a Gaussian function perpendicular to the oscillating structure at each time slice. The Gaussian fitting provides the centre of the oscillating structure at a particular time. To obtain the parameter of oscillations, we fit these centres using a sinusoidal function with a linear trend.

\begin{equation}
\label{eq1}
\centering
    \xi(t) = \xi_{0}+\xi_{1} \sin(2\pi t/P+\phi)+\xi_{2}t.
\end{equation}
where $\xi_{1}$ represents the oscillation amplitude, $P$ is the oscillation period, $\phi$ is the phase, and $\xi_{0}$ and $\xi_{2}$ are constants. We implemented the same techniques for calculating oscillation parameters in the 42 oscillations, and 6 are shown in Figure ~\ref{fig:xt-maps}. The best-fit functions are shown as red curves. {The error bars shown in Figure ~\ref{fig:xt-maps} are the standard errors on the centre position of the loop, obtained after fitting the Gaussian at each time slice.} The amplitude and period obtained after fitting, along with the values of propagated errors, are indicated in the $x-t$ maps. We did not use any automatic method to fit the oscillations as performed in previous statistical studies \citep{2014Thurgood, 2018Weberg, 2020Weberg}. The velocity amplitude of these oscillations was estimated using the relation $V = 2\pi\xi_{1}/P$. We computed the  uncertainty in velocity amplitude by $\sigma_{V}^2 = \left(\frac{\partial V}{\partial P}\sigma_{P} \right)^{2} + \left(\frac{\partial V}{\partial \xi_{1}}\sigma_{\xi_{1}} \right)^{2}$. The loop length was approximated by measuring the distance between footpoints, assuming a semicircular loop model. We calculated the loop length using the relation $L = \pi D/2$, where D is the distance between the loop footpoints. We assumed the uncertainty in the loop lengths to be 40\% (discussed in section \ref{sec_cor_seismo}).


\section{ Results and discussion}\label{results}

\subsection{Distribution of the parameters} \label{sect:dist_par}

\begin{figure*}[!ht]
\centering
\includegraphics[width=0.8\textwidth,clip,trim=0cm 0.8cm 0cm 1cm]{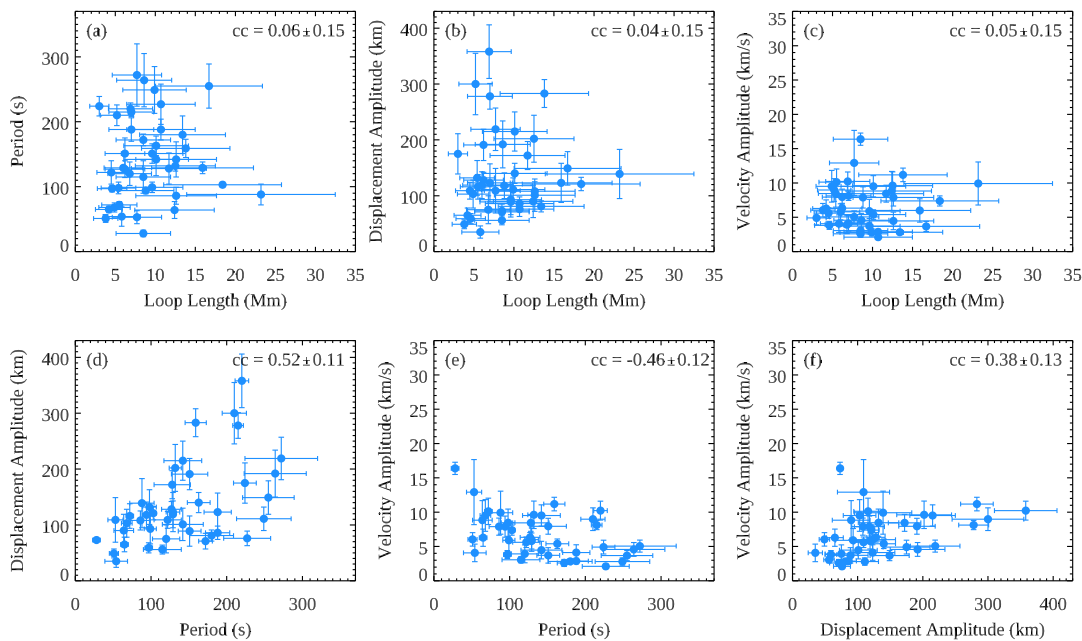}
\caption{Relations between parameters. Panels (a)-(f) show the scatter plots between the four estimated parameters loop length, period, displacement amplitude, and velocity amplitude. The linear Pearson correlation coefficient, along with standard errors, is also provided in each panel.}
\label{fig:correlations}
\end{figure*}
We estimated the oscillation amplitude, $\xi_{1}$, the period, $P$, the loop length, $L$, and the velocity amplitude $V$ for oscillations captured using $x-t$ maps and plotted their distribution. Figure ~\ref{fig:dist_param} shows the histograms of these parameters. The loop lengths vary from 3 to 23 Mm with a mean of 9.1$\pm$4.3 Mm. Previous studies of decayless oscillations in active-region loops estimated loop lengths of a few hundred megameters \citep{2013Anfinogentov, 2015Anfinogentov}. The study of kink oscillations in CBPs estimated a loop length in the range of 14-42 Mm \citep{2022Gao}. The loop length of the few loops we estimated here overlaps with the recent length measurement of active and quiet region loops using EUI \citep{2022Petrova, 2022Dong}. The average loop length in this analysis is shorter than in previous statistical studies using AIA because the resolution of EUI is high. The oscillation period lies between 28 to 272s, with a mean of 140$\pm$64s. Around 90\% of the oscillations have periods greater than 60s, which is similar to previously estimated periods for coronal loops \citep{2015Anfinogentov, 2019Nechaeva, 2022Gao, 2022Zhong}. Ten percent of the oscillation periods range from 28 to 60s, analogously to the decayless oscillations detected in small loops  \citep{2022Petrova, 2022Dong}. The displacement amplitudes are in the range of 35 to 358 km, with an average value of 134$\pm$72 km.  The average displacement amplitude is comparable to the estimated oscillation amplitude in active region coronal loops \citep{ 2015Anfinogentov, 2022Dong}. In contrast, it is larger than the estimated amplitudes for CBPs \citep{2022Gao}. When we consider a similar amount of energy flux in both scenarios, a greater displacement amplitude is required in smaller loops. The velocity amplitudes are between 2.1 to 16.4 km s$^{-1}$, with a mean of 6.6$\pm$3.1 km s$^{-1}$. The velocity amplitudes found in oscillations in CBPs, fine structures in active region moss, and in large-scale coronal loops are lower than 10 km s$^{-1}$ \citep{2013Anfinogentov,2013Morton, 2014Morton, 2016Nakariakov, 2022Gao}, whereas $\sim15\%$ of the oscillations in our study have velocity amplitudes larger than 10 km s$^{-1}$.

The mean period, amplitude, and velocity amplitude calculated in the statistical analysis of propagating kink waves in coronal plumes using AIA are greater than the estimated average of these quantities in the current work \citep{2014Thurgood, 2015Morton_natcom, 2018Weberg, 2020Weberg}. \cite{2019Morton} examined propagating waves in the fine-scale structure within the quiet Sun using EUV images. The mean of the period and velocity amplitudes are larger than those computed in this work. Furthermore, the velocity amplitudes of propagating waves were estimated by \cite{2011McIntosh} to be in the range of 20-25 km s$^{-1}$ in quiet regions and coronal holes, which is higher than the velocity amplitudes measured in this study.

\subsection{Correlation between different parameters}\label{sec:corre}

Figure ~\ref{fig:correlations} shows the scatter plots between the estimated parameters, described in section ~\ref{analysis}. The linear Pearson correlation coefficients, along with their standard errors, are also provided in the plots. The kink speed of an oscillating coronal loop will be proportional to loop length (L) and inverse of the period (P) \citep{Edwin&Roberts}. For a close range of kink speeds, the period is therefore approximately proportional to the loop length, as observed in the analysis of decayless oscillations of active region coronal loops with lengths of several hundred megameters \citep{2015Anfinogentov}. 

We found no significant correlation between the loop length and the period of oscillations (Figure \ref{fig:correlations}a). \cite{2022Gao} did not observe a significant correlation between the loop length and period in their study and suggested that these oscillations could be driven at the footpoints and can have periods similar to the driver, which can also be a possible reason for the poor correlation in our study. Recently, \citet{2022Dong} observed short-period oscillations in coronal loops with a mean length of $\sim15$ Mm. They found a strong correlation (0.98) between the loop length and the period.

 We found no correlation of the loop length with the displacement and velocity amplitudes. Moreover, our study indicated a correlation of 0.52 between the displacement amplitude and the period, as shown in the lower left panel of Figure ~\ref{fig:correlations}. It is possible that the correlation between the displacement amplitude and the period also depends on the sample and is therefore a selection effect. There may not necessarily be a physical relation between these two parameters. We find a negative correlation between the velocity amplitude and the period, as reported in \cite{2022Gao}. The positive correlation between the displacement and the velocity amplitude and the negative correlation between the period and the velocity amplitude might arise because they are directly related.

\begin{figure}[!ht]
\centering
\includegraphics[width=0.49\textwidth,clip,trim=2cm 2cm 2.5cm 2.5cm]{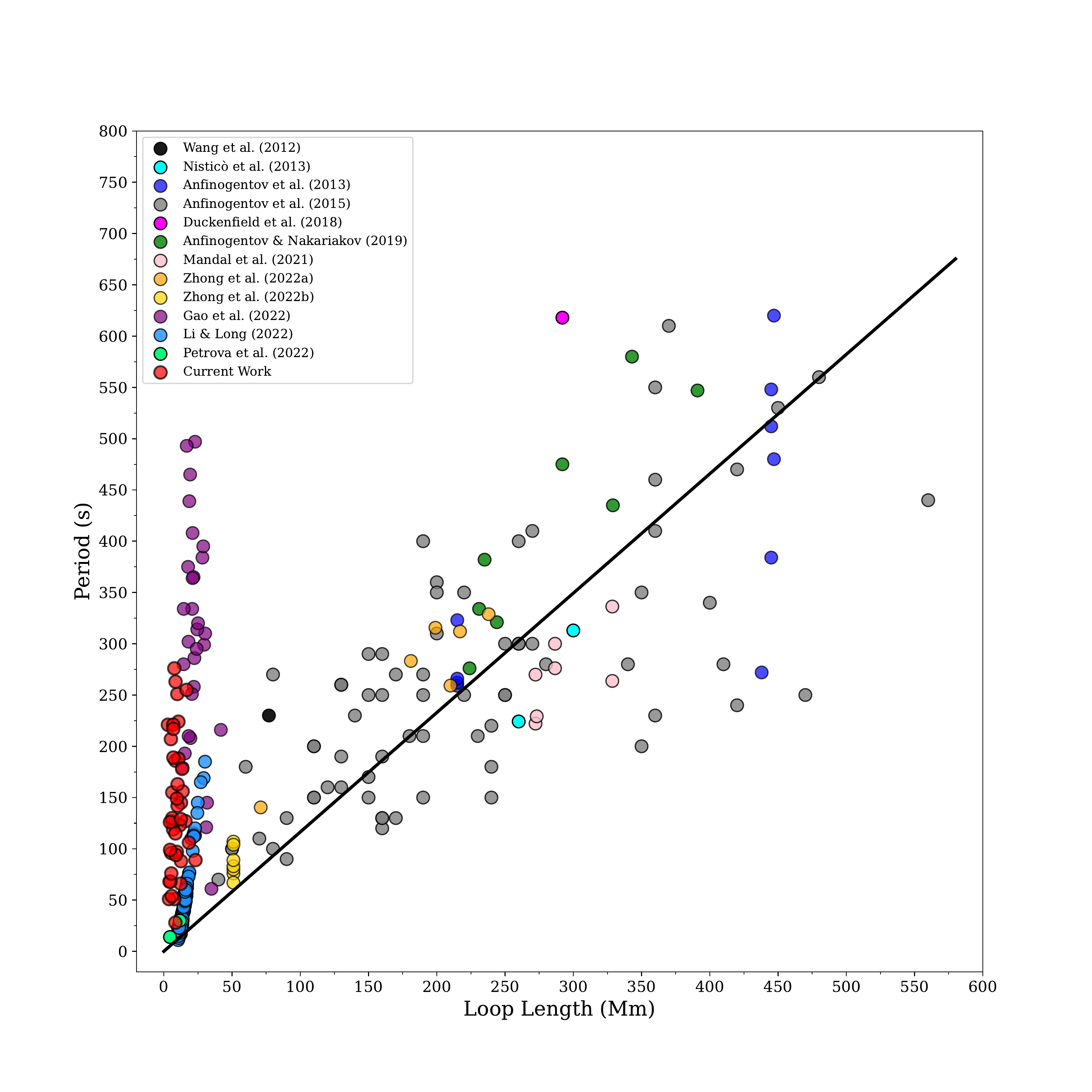}
\caption{Scaling between the loop length and the period. The figure shows the variation in the loop length vs. the period of the decayless oscillations analysed in previous studies \citep{2012Wang, 2013Nistic, 2013Anfinogentov, 2015Anfinogentov, 2018Duckenfield, 2019Anfinogento, 2021Mandal_flareoscl, 2022Zhonga, 2022Petrova, 2022Zhong, 2022Gao, 2022Dong}. The red points represent the results of the current work. The black lines represent the best fit for the datasets.}
\label{fig:all_studies}
\end{figure}

\begin{figure}[!ht]
\centering
\includegraphics[width=0.49\textwidth,clip,trim=3.1cm 0.8cm 3.6cm 2.1cm]{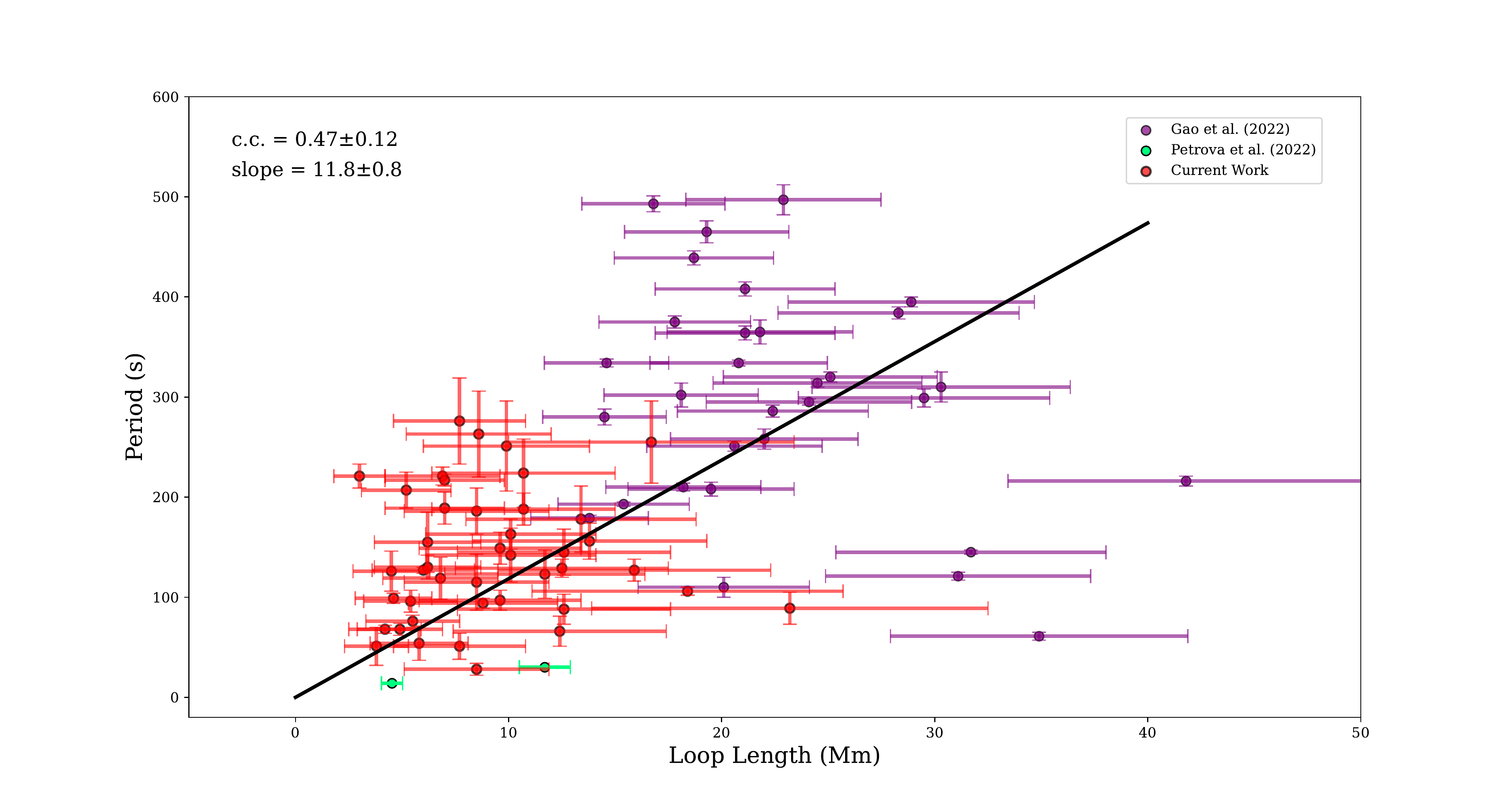}
\caption{Variation in the loop length vs. period for the small loops embedded in the quiet Sun and coronal holes. This figure displays the relation of the loop length and the period of the current work combined with \cite{2022Gao}, and \cite{2022Petrova}. The cross-correlation coefficients combining the studies are provided. The black lines represent the best fit for the datasets.}
\label{fig:gao_petrova_arpit}
\end{figure}

Figure \ref{fig:all_studies} presents the variation in the period with the loop length in our work, combined with various studies of decayless oscillations compiled from \citet{2012Wang, 2013Nistic, 2013Anfinogentov, 2015Anfinogentov, 2018Duckenfield, 2019Anfinogento, 2021Mandal_flareoscl, 2022Zhonga, 2022Petrova, 2022Zhong, 2022Gao, 2022Dong}. The black line provides the least-squares fit to these data points with a slope of 1.16$\pm$0.05. The uncertainty in the measurement of the loop length \citep[see][]{2007Tom,2021Berghmans} might affect the correlation between the loop length and the period in our study. However, it is unlikely to cause a significant shift in the direction of the length axis for the current work, preventing them from being aligned with the L versus P relation seen for longer loops (see Figure \ref{fig:all_studies}).

{Figure \ref{fig:gao_petrova_arpit} shows the variation in the loop length and the period only for loops found in the quiet Sun and coronal holes that were studied in \cite{2022Gao}, \cite{2022Petrova}, and this work. We assumed an uncertainty of 20\% in the loop length estimation for loops studied in \cite{2022Gao}. The red data points were obtained in this study, which indicates that the average loop length is shorter than in previous statistical studies. The dependence of the oscillation period on the loop length suggests that the shorter loops will have short periods \citep{2015Anfinogentov, 2016Goddard, 2022Zhong, 2022Dong}. However, several long-period oscillations are observed in CBPs with an average loop length of $\sim$23 Mm \citep{2022Gao}.  We also find many oscillations with longer periods in loops with a mean length of $\sim$10 Mm (see Figure \ref{fig:all_studies}). We find almost no correlation between the loop length and the period obtained in this work. However, when we consider loops embedded in the quiet Sun and coronal holes, the cross-correlation coefficient is improved to be 0.47$\pm$0.12, which is greater and more significant than the correlation coefficient obtained in the present study. It should be noted that the properties of loops that are embedded in the quiet Sun and coronal holes can be different because they form in different magnetic field configurations.  However, the difference between the populations of small loops in the quiet Sun and coronal holes and its effect on the oscillations is not yet clear.}

\begin{table*}[]
\centering
\caption{Details of the oscillating loop and oscillation parameters}
\begin{tabular}{@{}ccccccccc@{}}
\toprule
{No.} & {Date} & {X (arcsec)} & {Y (arcsec)} & {Start Time (UT)} & {End Time (UT)} & {L (Mm)} & {P (s)} & $\boldsymbol{\xi_{1}}$ {(km)} \\ \midrule
1  & 2021-09-14 & -101.1 & -1552.2 & 05:59:42 & 06:05:32 & 9.9  & 249 ± 36 & 111 ± 21 \\ \midrule
2  & 2021-09-14 & -304.1 & -1535.7 & 05:57:12 & 06:01:47 & 15.9 & 129 ± 9  & 123 ± 35 \\ \midrule
3  & 2021-09-14 & -158.7 & -1524.2 & 05:53:02 & 05:56:22 & 11.7 & 128 ± 24 & 172 ± 25 \\ \midrule
4  & 2021-09-14 & -101.4 & -1552.3 & 06:02:37 & 06:09:42 & 8.5  & 172 ± 9  & 71 ± 14  \\ \midrule
5  & 2021-09-14 & -111.5 & -1327.7 & 05:58:52 & 06:01:22 & 12.4 & 64 ± 13  & 90 ± 29  \\ \midrule
6  & 2021-09-14 & 163.1  & -1019.1 & 05:53:02 & 05:58:52 & 16.7 & 255 ± 34 & 149 ± 30 \\ \midrule
7  & 2021-09-14 & 192.2  & -1478.6 & 05:54:42 & 06:03:02 & 10.7 & 227 ± 31 & 76 ± 13  \\ \midrule
8  & 2021-09-14 & 493.7  & -1264.2 & 05:53:02 & 05:57:12 & 6.2  & 151 ± 24 & 191 ± 28 \\ \midrule
9  & 2021-09-14 & 488.3  & -1251.2 & 05:53:02 & 05:58:02 & 10.7 & 188 ± 16 & 86 ± 12  \\ \midrule
10 & 2021-09-14 & 392.7  & -1014.7 & 05:53:52 & 06:02:12 & 5.2  & 210 ± 16 & 300 ± 55 \\ \midrule
11 & 2021-09-14 & 113.1  & -1276.1 & 05:56:22 & 05:59:42 & 12.6 & 142 ± 27 & 101 ± 29 \\ \midrule
12 & 2021-09-14 & 291.3  & -1158.6 & 06:03:02 & 06:11:22 & 8.6  & 264 ± 41 & 192 ± 42 \\ \midrule
13 & 2021-09-14 & 114.4  & -1275.5 & 05:57:02 & 05:59:42 & 12.6 & 86 ± 13  & 108 ± 17 \\ \midrule
14 & 2021-09-14 & 140.7  & -1605.2 & 06:06:22 & 06:08:37 & 5.4  & 98 ± 9   & 132 ± 31 \\ \midrule
15 & 2021-09-14 & 288.2  & -1548.4 & 06:06:12 & 06:11:37 & 7.7  & 272 ± 48 & 219 ± 38 \\ \midrule
16 & 2021-09-14 & 96.1   & -1592.2 & 06:02:17 & 06:05:47 & 10.1 & 142 ± 25 & 215 ± 35 \\ \midrule
17 & 2021-09-14 & -763.3 & 1287.8  & 04:16:27 & 04:19:22 & 23.2 & 88 ± 16  & 139 ± 44 \\ \midrule
18 & 2021-09-14 & -314.1 & 1317.8  & 04:13:32 & 04:16:47 & 6.8  & 120 ± 18 & 75 ± 25  \\ \midrule
19 & 2021-09-14 & -320.6 & 1335.7  & 04:21:27 & 04:26:37 & 13.8 & 159 ± 14 & 283 ± 25 \\ \midrule
20 & 2021-09-14 & -514.3 & 1562.5  & 04:13:07 & 04:15:12 & 7.7  & 53 ± 10  & 109 ± 40 \\ \midrule
21 & 2021-09-14 & -8.1   & 1557.7  & 04:11:27 & 04:15:12 & 9.6  & 99 ± 8   & 93 ± 27  \\ \midrule
22 & 2022-03-30 & -150.1 & -2791.6 & 04:35:42 & 04:41:30 & 8.8  & 94 ± 5   & 118 ± 23 \\ \midrule
23 & 2022-03-30 & -270.1 & -2416.1 & 04:31:15 & 04:33:09 & 3.8  & 51 ± 6   & 49 ± 8   \\ \midrule
24 & 2022-03-30 & -613.3 & -2562.2 & 04:41:39 & 04:51:54 & 3.0  & 224 ± 15 & 175 ± 36 \\ \midrule
25 & 2022-03-30 & -612.3 & -2557.8 & 04:31:06 & 04:37:54 & 6.9  & 220 ± 9  & 358 ± 48 \\ \midrule
26 & 2022-03-30 & -584.1 & -2714.7 & 04:41:30 & 04:45:15 & 4.6  & 97 ± 5   & 59 ± 9   \\ \midrule
27 & 2022-03-30 & -303.5 & -2626.8 & 04:31:36 & 04:35:00 & 8.5  & 115 ± 25 & 56 ± 8   \\ \midrule
28 & 2022-03-30 & -303.3 & -2625.8 & 04:40:00 & 04:44:33 & 9.6  & 151 ± 15 & 89 ± 27  \\ \midrule
29 & 2022-03-30 & -220.0 & -2532.9 & 04:31:36 & 04:37:30 & 7.0  & 188 ± 18 & 123 ± 34 \\ \midrule
30 & 2022-03-30 & 205.4  & -2643.5 & 04:55:15 & 04:56:24 & 8.5  & 28 ± 5   & 73 ± 4   \\ \midrule
31 & 2022-03-30 & 247.5  & -2313.7 & 04:42:30 & 04:51:30 & 6.0  & 129 ± 3  & 119 ± 24 \\ \midrule
32 & 2022-03-30 & 269.4  & -2324.2 & 04:32:00 & 04:38:00 & 6.2  & 127 ± 13 & 128 ± 14 \\ \midrule
33 & 2022-03-30 & 253.7  & -2316.7 & 04:30:00 & 04:32:30 & 4.9  & 68 ± 7   & 103 ± 25 \\ \midrule
34 & 2022-03-30 & 231.3  & -2306.0 & 04:32:00 & 04:35:30 & 4.5  & 122 ± 18 & 109 ± 15 \\ \midrule
35 & 2022-03-30 & 260.9  & -2320.1 & 04:50:00 & 04:55:30 & 4.2  & 65 ± 5   & 65 ± 13  \\ \midrule
36 & 2022-03-30 & 71.4   & -2275.9 & 04:31:06 & 04:33:42 & 5.8  & 54 ± 15  & 35 ± 11  \\ \midrule
37 & 2022-03-30 & 162.7  & -2864.4 & 04:43:00 & 04:48:54 & 12.5 & 132 ± 8  & 202 ± 42 \\ \midrule
38 & 2022-03-30 & 153.7  & -2856.3 & 04:53:00 & 04:59:30 & 7.0  & 215 ± 7  & 278 ± 23 \\ \midrule
39 & 2022-03-30 & 155.1  & -2855.6 & 04:32:45 & 04:38:15 & 10.1 & 163 ± 15 & 140 ± 18 \\ \midrule
40 & 2022-03-30 & -105.6 & -2829.7 & 04:30:54 & 04:35:12 & 13.4 & 180 ± 29 & 81 ± 10  \\ \midrule
41 & 2022-03-30 & 169.0  & -2863.0 & 04:45:15 & 04:49:15 & 18.4 & 103 ± 4  & 121 ± 12 \\ \midrule
42 & 2022-03-30 & -62.4  & -2917.0 & 04:54:30 & 04:58:00 & 5.5  & 72 ± 5   & 116 ± 22 \\ \bottomrule
\end{tabular}
\end{table*}

\begin{figure}[!ht]
\centering
\includegraphics[width=0.49\textwidth,clip,trim=0cm 0cm 0cm 0cm]{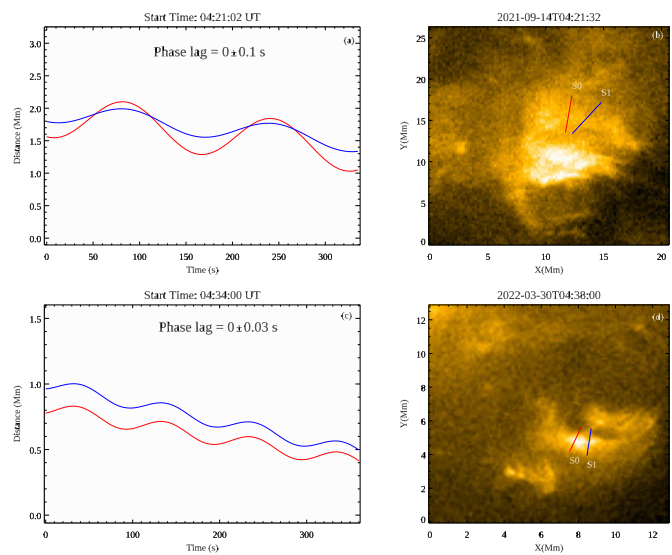}
\caption{{Phase lag analysis of the oscillations. The figure presents the multi-slit analysis for the two loops shown in panels (b) and (d). S0 and S1 are two slits that were placed at different positions of the loops. Panels (a) and (c) show the fitted oscillation profile from slits S0 and S1 indicated in panels (b) and (d). The oscillations detected in S0 and S1 show no significant phase shift.} }
\label{fig:wave_mode}
\end{figure}

\begin{figure*}[!ht]
\centering
\includegraphics[width=0.8\textwidth,clip,trim=0.2cm 3cm 0.5cm 0.5cm]{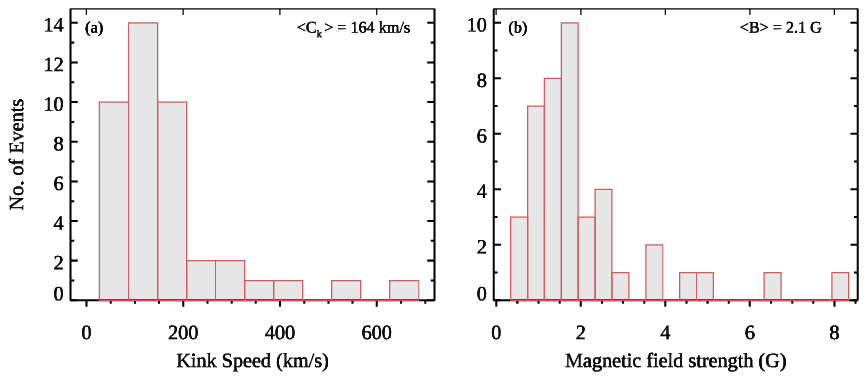}
\caption{Results from coronal seismology. The figure shows the histograms of the kink speeds, C$_{k}$, and magnetic field strength, B, obtained using the estimated oscillation parameters. The mean value of the distribution is also provided.  }
\label{fig:seismology}
\end{figure*}

\subsection{{Phase lag analysis}}\label{sec_phase_lag}

To understand whether the observed waves are standing or propagating, we positioned multiple slits along each loop in the sample. We examined the oscillation detected at different slits to capture any significant phase shift. These loops are characterized by their small size, short duration, and highly dynamic nature. Due to the dynamic behaviour of these loops, it is challenging to accurately place the slits and identify a reliable oscillation pattern at the footpoints. Additionally, the loops are not clearly distinguishable from the background at different locations along their length, making it difficult to capture oscillations in each part of the loop. We find that approximately 50\% of the loops did not exhibit any noticeable phase shift in the oscillations captured at different positions of the loop. For the remaining loops, no distinct oscillation signatures were detected at slits far from the apex. Nevertheless, no phase shift was obtained close to the apex of these loops.
                                                                        
{Figure \ref{fig:wave_mode} shows two loops along with the oscillation signatures at two artificial slits. Figure \ref{fig:wave_mode}(a) and (c) depicts the oscillations profiles at slit positions S0 (red) and S1 (blue), shown in Figure \ref{fig:wave_mode}(b) and (d). These oscillation profiles were obtained after fitting the $x-t$ maps of these slits. The $x-t$ maps used for phase analysis are unsharp masked to minimise the effect of the variable background and to better detect the oscillations. S1 (Figure \ref{fig:wave_mode}(b)) is situated between the loop top and right footpoint, and the amplitude of the oscillation for S1 is smaller than the amplitude in S0. Figure \ref{fig:wave_mode}(a) and (c) shows that no significant phase shift is present between oscillations at S0 and S1. The phase lags for these oscillation time-series are 0$\pm$0.1 s and 0$\pm$0.03 s. The uncertainty in the phase lag is calculated by fitting a Gaussian profile near the maximum phase lag.   }

We detected transverse oscillations lasting for $\sim$ 1.5-5 cycles at multiple slits at different loop positions. If the loops are not stable for a longer duration, then oscillations cannot be detected for several cycles. Thirty-two ($\sim$75\%) of the detected oscillations were found to occur with cycles equal to or greater than two. It is not possible to identify the decay in amplitude for oscillations with fewer than two cycles because the decay time of the decaying oscillations is mostly greater than two cycles. However, oscillations with cycles equal to or greater than two exhibit no apparent decay in amplitude.

\subsection{{Possible wave modes}} \label{sec_possible_wave_mode}

\subsubsection{{Standing wave}}
{Unlike the decayless oscillations observed in larger loops, which demonstrate a linear relation between the loop length and the period, which is indicative of a standing wave mode \citep{2013Anfinogentov, 2015Anfinogentov}, our study revealed no strong correlation between the loop length and the period. This suggests that the nature of these waves may differ from the expected standing-wave behaviour. We focused our attention exclusively on small loops located in the quiet Sun and coronal holes, as depicted in  Figure \ref{fig:gao_petrova_arpit}, which include \citet{2022Gao}, \cite{2022Petrova}, and this study. In this subset, we found a correlation coefficient of 0.47 between the loop length and the period, which is notably higher than the correlation observed in this study alone. However, even with this improvement, the correlation is still insufficient to firmly support the interpretation of a standing-wave behaviour.}

 {In addition, in limited studies of coronal loops, the oscillations measured at different slit positions showed an in-phase behaviour, indicating a standing-mode interpretation \citep{2015Anfinogentov}.  Nonetheless, when we consider the phase lag obtained in this study, they may be indicating a standing-wave interpretation. However, as the loop length and period showed no strong correlation, it is not possible to conclude that these waves are standing.  }

\subsubsection{{Propagating wave}}

{The absence of a significant correlation between the loop length and the period in this study can be considered to mean a wave mode other than a standing wave. A scenario of a fast-propagating wave appears to be a possibility. When we assume that the speeds of fast-propagating waves range from 500-600 km/s and the imaging cadences are 3s and 5s (Table 1), then the slits need to be at least several megameters apart to detect the propagation of waves. Because the loop half-length (apex to footpoint) is of comparable size, it will in most cases, be hard to identify the propagating waves. Because it takes a considerable amount of time for a wave to traverse the entire length of large stable loops, the measurement of the phase shift between oscillations at various slit positions is not restricted by the observational cadence of EUV imagers such as AIA. It is difficult to eliminate the possibility of propagating waves completely, but it is equally challenging to prove their existence in the current analysis. Consequently, it is not feasible to assert with absolute certainty whether these waves are standing or propagating.}

\subsection{Coronal seismology and the estimation of the energy flux}\label{sec_cor_seismo}

{As discussed in the previous section, the wave mode of these waves is uncertain, and a standing nature is a possibility. The calculations and estimates provided in this section hold true under the assumption that the oscillations are indeed standing.}

For a standing wave, the kink speed (C$_k$) can be calculated as
\begin{equation*}
    C_{k} = \frac{2L}{P}.
\end{equation*}
We assumed that the observed oscillations are fundamental modes. Figure \ref{fig:seismology}(a) shows the distribution of the kink speed. The kink speed has a range of 27-630 km s$^{-1}$ with an average of 164$\pm$123 km s$^{-1}$. The estimated kink speed for most loops is lower than that of active region loops \citep{2019Anfinogento}, but comparable to the kink speeds in CBPs. Using the internal and external intensities ratio, \cite{2022Gao} showed that the internal Alfv\'en speed is correlated to the loop length with a correlation coefficient of 0.63. Although we did not estimate the internal and external intensities, the kink speed and loop length show a correlation of 0.55 in our study. This supports the idea that the Alfv\'en speed increases with height and is higher for larger loops.    

The magnetic field strength, B, can be estimated using the following relation:
\begin{equation*}
    B = C_{k}\sqrt{\mu_{0}\rho_{i}\widetilde{m}}\sqrt{\frac{1+\rho_{e}/\rho_{i}}{2}}, 
\end{equation*}
where $\rho_{i}$ and $\rho_{e}$ are the internal and external plasma density, and $\mu_{0}$ and $\widetilde{m}$ denote the magnetic permeability in vacuum and the mean molecular weight. We took $\rho_{i}$ = 10$^{9}$cm$^{-3}$ and a density contrast, $\rho_{e}/\rho_{i}$ = 1/3 to calculate the magnetic field \citep{2022Gao, 2022Petrova, 2022Dong}. Figure \ref{fig:seismology}(b) shows the distribution of estimated magnetic field strengths.  The average value of the magnetic field is 2.1$\pm$1.5 G, with a range of  0.34-8 G. These loops are embedded in the quiet Sun and coronal holes, and most of the loops have a magnetic field strength lower than 4 G. These estimates of the magnetic fields are lower than previous estimates using seismology for loop lengths of a few hundred megameters in an active region \citep{2013Nistic}. The potential field extrapolation of photospheric magnetograms for x-ray bright points shows that several loops with a length of $\sim$ 10 Mm have a magnetic field strength greater than 10 G \citep{2023Mondal}. However, the average value of the magnetic field obtained here is similar to quiet-Sun magnetic fields calculated using magnetoseismology of propagating kink waves \citep{2020Yang}. \cite{2011West} found a magnetic field strength of 0.7$\pm$0.7 G in the quiet-Sun region by performing coronal seismology using EIT waves. The underestimation of the loop length could result in a lower kink speed. The underestimation of the loop length can be about 10\% for loops of a few hundred megameters \citep{2000Mackay, 2007Tom}. The analysis of campfire loops with a loop length of $\sim$3-4 Mm indicated that this underestimation could increase up to $\sim$ 40-50\%  as it is possible that the part of the loop that is embedded in the chromosphere and photosphere is not observed in 174 \AA\ emission \citep{2021Berghmans}. It must be noted that even when we add a 50\% error in length, the average value of the kink speed would be 245 km s$^{-1}$, which remains smaller than those observed in the coronal heights. Because the kink speed depends on the internal and external Alfv\'en speeds, it might be possible that these structures have quite different density contrasts and/or magnetic fields compared to large-scale coronal loops. 

The energy flux carried by the observed kink waves can be estimated  by the following expression \citep{2014Van, 2022Petrova}:
\begin{equation*}
    E_{f} = \frac{1}{2} C_{k} (\rho_{e}+\rho_{i}) V^{2} .
\end{equation*}
Using the distribution of C$_{k}$ and V, and considering the values of the internal plasma density and density contrast (mentioned in section \ref{sec_cor_seismo}), we find a broad range of the energy flux, 0.6-313 W m$^{-2}$. The mean and median of the energy flux distribution are 19 and 6 W m$^{-2}$, respectively. The energy flux of 313 W m$^{-2}$ is obtained for the period of 28 s with a displacement amplitude of $\sim$ 73 km. This suggests that short-period oscillations have a high energy flux in the quiet Sun and coronal holes.  \cite{2022Petrova} analysed two short-period oscillations of $\sim$ 14 and 30 s in the quiet-Sun region. The wave energy fluxes calculated for these oscillations were 1.9 and 6.5 kW m$^{-2}$, but in this work, we find much lower values for the energy flux.  This potentially means that high-energy flux events such as in \cite{2022Petrova} are not so common in the quiet Sun. We thoroughly examined each small-scale loop in the datasets and observed these oscillations on only a few occasions. Within the limited duration of the observation and the limited field of view of data, we were unable to find the oscillations to be omnipresent like for the propagating waves in the CoMP. \cite{2022Gao} presented long-period low-energy decayless waves in bright points. In our study, we find both short-period and long-period oscillations. Figure \ref{fig:all_studies} shows that our study has revealed periods and loop lengths that were absent in \cite{2022Gao} and \cite{2022Petrova}.  Although the maximum energy flux we estimated is sufficient to heat the quiet corona \citep{1977Withbroe}, the mean and median energy fluxes indicate that decayless waves with high-energy fluxes are not prevalent in the quiet Sun and coronal holes.

\section{Summary and conclusion}\label{conclusion}
We performed the statistical analysis of decayless kink oscillations in small loops using high-resolution observation from the HRI$_{\rm{EUV}}$ telescopes of EUI. The oscillating loops have lengths of about 10 Mm, and they are dynamic in nature. The average loop length in this study is shorter than in previous statistical studies of decayless kink oscillations in coronal loops. The analysed loops have a mean oscillation period and displacement amplitude of 140s and 134 km, respectively.  {The analysis we carried out to identify wave modes suggests the potential presence of standing waves. Nonetheless, the identification of propagating waves with the datasets used here proves to be a challenging endeavour. Moreover, the absence of a significant correlation between the loop length and period precludes their categorisation as standing-wave modes. The available evidence currently does not suffice to draw a definitive conclusion of whether these waves are standing or propagating.} Further insights into the nature of these waves can be gained through numerical simulations of transverse oscillations in small loops and additional studies focusing on the same phenomenon.  { The kink speed, magnetic field, and energy flux of these waves were calculated considering that standing waves are one possibility.} The estimation of the kink speed shows a range of $\sim$27-630 km s$^{-1}$, which is lower than the kink speed found in loops of several hundred megameters in active regions. We estimated the magnetic field strength, which indicated lower values than were obtained in active region loops.  The energy flux estimation provided a range of $\sim$0.6-313 W m$^{-2}$. We find that the energy flux of most oscillations is insufficient to compensate for the radiative losses in the quiet corona and coronal holes.  This indicates that transverse oscillations with high-energy flux in the quiet Sun and coronal holes may not be prevalent. {In conclusion, if the waves identified in this study are standing, it implies that the transverse waves in small loops, immersed in coronal holes and quiet corona, cannot provide significant energy to balance the radiative losses and accelerate the solar wind.} 

\begin{acknowledgements}
A.K.S is supported by funds of the Council of Scientific \& Industrial Research (CSIR), India, under file no. 09/079(2872)/2021-EMR-I.  V.P. is supported by SERB start-up research grant (File no. SRG/2022/001687). A.K.S. acknowledges Krishna Prasad Sayamanthula and Ritesh Patel for their thoughtful discussions. A. N. Z. thanks the Belgian Federal Science Policy Office (BELSPO) for the provision of financial support in the framework of the PRODEX Programme of the European Space Agency (ESA) under contract number 4000136424. T.V.D. was supported by the European Research Council (ERC) under the European Union's Horizon 2020 research and innovation programme (grant agreement No 724326), the C1 grant TRACEspace of Internal Funds KU Leuven, and a Senior Research Project (G088021N) of the FWO Vlaanderen. Solar Orbiter is a space mission of international collaboration between ESA and NASA, operated by ESA. The EUI instrument was built by CSL, IAS, MPS, MSSL/UCL, PMOD/WRC, ROB, LCF/IO with funding from the Belgian Federal Science Policy Office (BELSPO/PRODEX PEA 4000112292 and 4000134088); the Centre National d’Etudes Spatiales (CNES); the UK Space Agency (UKSA); the Bundesministerium f\"{u}r Wirtschaft und Energie (BMWi) through the Deutsches Zentrum f\"{u}r Luft- und Raumfahrt (DLR); and the Swiss Space Office (SSO). 
\end{acknowledgements}

\bibliographystyle{aa}
\bibliography{kink_polar}

\end{document}